# EFFICIENT TIME SYNCHRONIZED ONE-TIME PASSWORD SCHEME TO PROVIDE SECURE WAKE-UP AUTHENTICATION ON WIRELESS SENSOR NETWORKS

Salem Aljareh[1] and Anastasios Kavoukis[2]

[1]School of Engineering, University of Portsmouth, Portsmouth, United Kingdom
`Salem.Aljareh@port.ac.uk`
[2]School of Engineering, University of Portsmouth, Portsmouth, United Kingdom
`anastasios.kavoukis@port.ac.uk`

**ABSTRACT**

*In this paper we propose Time Synchronized One-Time-Password scheme to provide secure wake up authentication. The main constraint of wireless sensor networks is their limited power resource that prevents us from using radio transmission over the network to transfer the passwords. On the other hand computation power consumption is insignificant when compared to the costs associated with the power needed for transmitting the right set of keys. In addition to prevent adversaries from reading and following the timeline of the network, we propose to encrypt the tokens using symmetric encryption to prevent replay attacks.*

**KEYWORDS**

*Wake-up, Security, Wireless Sensor Networks, One-Time Password*

## 1. INTRODUCTION

On duty cycling protocols, nodes periodically activated to perform communications and sleep after all the tasks are completed to enable them to save power until the next scheduled wake up [1, 2]. The problem related to this protocol is that nodes will have to wake up even if communications are not necessary so they can monitor the channel for data; as a result the nodes waste precious energy resources to activate the receiver and the MCU (microcontroller unit) in order to listen to the channel. In addition depending on the application's requirements, duty cycle protocols are unable to transmit the data on demand to the channel because the transmissions are scheduled[3].

On the other hand asynchronous protocols do not face the same duty cycle problem[4] because the sleeping operation is not a scheduled procedure nodes can sleep most of the time until they are required to wake up and instantly transmit data when it is necessary. Since sleep is not scheduled in these protocols, it means that the sleep can be interrupted in several ways which causes nodes to become vulnerable to malicious attacks such as denial of sleep[5]. Denial of sleep attack can either interrupt the sleep or prevent the node from going to sleep after transmission which results in an unnecessary drain on the power resources of the node.

Another problem that asynchronous protocols face is the authentication of requests. The physical characteristics of radio communications demand power in order to receive and calculate the integrity of a message. In cases where the captured data failed to authenticate the node will have to use power to reject it. In this paper we propose the use of tokens to initiate a connection before the establishment of communication in order to avoid any unnecessary power consumption on rejected messages.

The rest of the paper is organized as follows. Section 2 will analyse previous work undertaken by other researchers. Section 3 will evaluate the use of wakeup receivers and one time passwords to counter the 'Denial of Sleep' attack.

## 2. RELATED WORK

The proposed scheme benefit's from two previous areas of research. Firstly "wake up receivers" that investigates the use of an additional, very low power receiver, to be used only for the purpose of receiving an asynchronously signal and wake up the rest parts of the node as and when required. Secondly "One-Time-Password authentication" that is used to generate passwords with a short validity period i.e.: one login or one transaction. Their purpose is to avoid replay attacks by potential intruders capturing the password.

### 2.1 Wake-up receivers

Wake-up receivers are mainly designed to be used on asynchronous communication models to receive only "request to send" signals[6]. Their physical characteristics allow them to monitor a radio channel and receive signals using negligible amounts of energy on low data rates. One of the first designs of wake-up receivers for wireless sensor networks designed to extract the power of the signal in order to provide energy to operate the receiver circuit[7]. Plethora of researches on Wake-up receivers, have also been conducted on passive RFID that provides them with viable solution as proposed on [2]. Wake-up scheme is vital for passive RFID devices because they rarely use their communications. One of the most recent papers [8]combines wake-up scheme with security and possible attacks along with solutions to secure using specific tokens as signals to wake up the procedure.

### 2.2 One-Time Password

Since 1981 when Lamport introduced one time password schemes, many banks authentication systems are now using his theory to prevent reuse of a static password [9]. The main idea of one-time password schemes is that the password changes on each authentication and derives either from a static mathematical expression or by the actual time of day and changes periodically which is called counter one-time-password or time synchronized one-time-password.

The merit of time synchronized protocols is that it does not require complex calculations or a certification authority but only a counter to maintain the synchronization. On the other hand to achieve tolerance, the actual time has to be separated into time slots and pair the valid passwords for each time slot. However the algorithm becomes ineffective in cases where multiple simultaneous authentications are attempted on a single time slot.

## 3. CRITICAL ANALYSIS OF PREVIOUS RESEARCH

This section covers a discussion for our analytical study for previous published researched contributed to the solution for denial of sleep attacks. The study includes simulation of the most related solution.

The most common transmission synchronization protocol for wireless sensor networks is synchronous transmission [3]. The reason researches prefer to use synchronous synchronization is because it allows nodes to sleep most of the time so they can save power resources and wake up after a pre-specified period of time to resume communications. This advantage however is coupled with a major problem. Nodes always have to wake up even if transmissions are not necessary in order to check for incoming data from a neighbour requesting to carry data to the sink (multi-hop environment). Equation 1 presents the amount of energy $E_{scheduled}$ that is consumed regardless whether or not it transmits data and derives from the following parameters.

Where $P_{switch}$ is the power required to switch from active mode to sleep mode or from sleep to active mode.

Where $t_{active}$ is the time that the node remains active.

Where $P_{active}$ is the power consumed while the node is on active state.

Where $P_{transceive}$ is the power consumed for transmission and reception.

$$E_{scheduled} = (P_{switch} * 2) + (t_{active} * P_{active}) + P_{trancieve}$$

Equation 1: Energy required for each scheduled wakeup up on synchronous protocols.

Asynchronous models on the other hand do not need to periodically wake up in order to check for incoming data. However they are considered to be highly exposed to denial of sleep attacks on insecure environments. This happens because nodes will have to accept wake up on any wake request either by other nodes or by an adversary that is sending malicious request to keep the node busy on purpose. Previous attempts to prevent the power exhaust of the batteries have been conducted using detection of anomaly methods[10]. The results of these researches are satisfactory enough regarding the prevention of power surge but only after the attack have succeeded several times to trigger the detection and also insufficient to proceed with a normal operation after the detection. So from above we consider intrusion detection systems incapable of counter attack a denial of sleep attack.

A viable solution we have identified fuelled us to investigate furthermore the use of a wake up receiver as proposed on [8]. Rainer Falk et al proposed the use of the wake up receiver as a door keeper that wakes up the rest of node's parts only when the wake up receiver receives a valid token. In spite of the contribution to the community and intellectual nourishment they offered us we consider the research incomplete in an energy aspect. The lack of deep investigation on token exchange leads us to consider that nodes will have to spend energy on transmitting next useable token or a key-chain to the network. In addition even if we consider that key chain is capable enough to cover the entire lifetime instead of token exchange for each node, we can also present a scenario that the scheme will fail. Figure 1 illustrates a simulation for a scenario we designed to visualize the vulnerability we have identified in Falk et al's proposed scheme. We have used three nodes communicating asynchronously.

1. All nodes on initialization stage they exchange their tokens so they can use them to remotely wake up each other.
2. Later we trigger Node 0 to transmit to node 1 while node 2 is sleeping and it succeeds.
3. Node1 wakes up and receives the data.
4. Both Node0 and Node1 go back to sleep.
5. Node1 will change the active token because it was exposed to the aerial and a possible eavesdropper could be hearing and capture the data.

6. But when later Node 2 decides to transmit to Node1, will attempt an authenticating using the same token that was used before by Node0.
7. Node1 will reject it because it is not the current active token to avoid reuse of the previous token by an adversary node.

As a result, nodes are incapable of following any key change during their sleep and will assume that their active remote token will be still available to be used but they will fail instead.

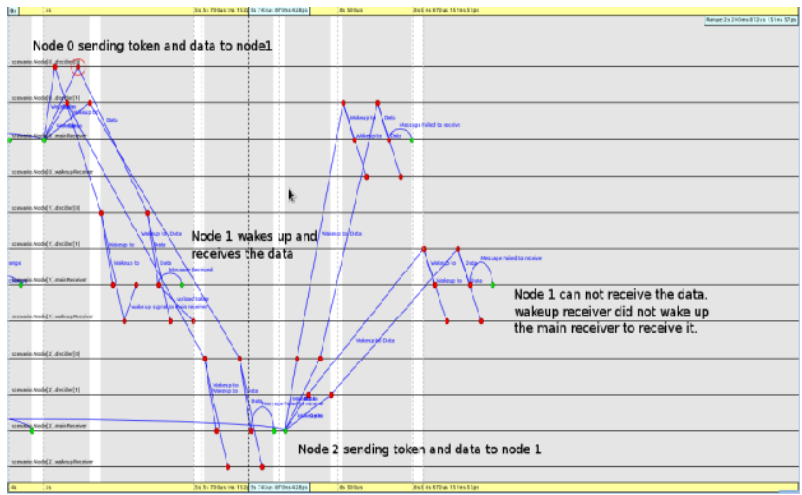

Figure 1: Vulnerability of previous research

## 4. PROPOSED SCHEME

The basis of our proposed protocol was to investigate an efficient and secure asynchronous wake-up scheme for wireless sensor nodes. The main idea was to keep the node always in sleep mode and wake it up only as and when communications were necessary. Therefore to enable us to accomplish an efficient but highly available protocol we have to use an additional receiver that is capable of remaining in idle mode utilising only negligible amounts of energy as shown in Figure 2. This receiver will be responsible for capturing communication requests and wake up the parts of the node which are currently at rest in order to receive the actual data.

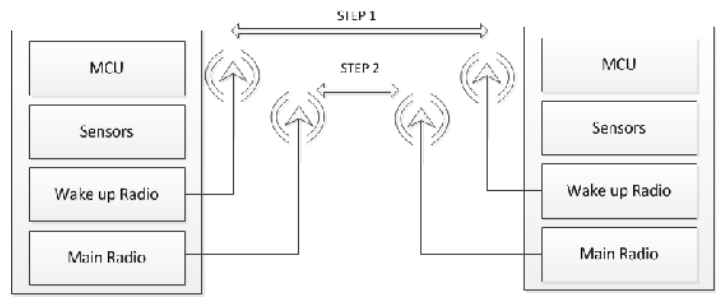

Figure 2: Step by step wake up procedure using node's schematic

The security consideration of this scenario is to authenticate the incoming requests to the wake up receiver on physical layer so that potential adversaries targeting to drain its energy by trying to wake up the node will fail before the main receiver receives the request and try to authenticate it using the MCU, in other words this scenario benefits from moving the authentication from application layer to physical layer. In order to ensure the security of the protocol, we need to consider all possible attack methods that an adversary could use whilst always keeping in mind that power costs are the most important factor. For example we cannot use public key generation on a sensor node because the calculations are too complex and would take a lot of time and energy to complete. Although numerous research exploring the possibilities of using public key encryption has been undertaken but there is a constraint on how frequently the public key can be generated thus making it inefficient [11].

The most obvious security threat that we have to tackle is a possible reuse of a wake-up token in the case that an adversary was listening to the channel when the token was transmitted and captured it (replay attack). Replaying this token would cause the sleeping node to reawaken causing a Denial of Sleep (DoS) attack [12, 13] as shown in Figure 3.Therefore we have to differentiate the token for each unique use thereby rendering the captured token useless should an attempt be made to use it surreptitiously.

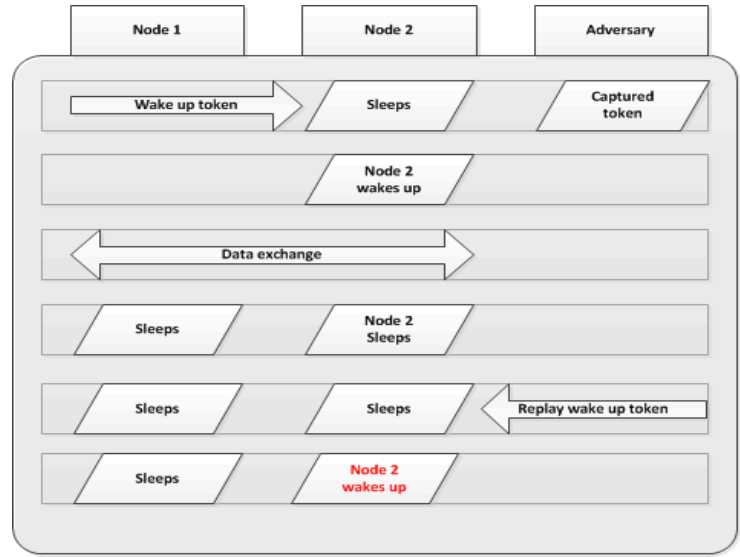

Figure 3: Schematic of our solution for remote wake up without exchanging tokens.

## 4.1 Token Generation

The request that wireless sensor nodes will use to securely wake up each other is called token. Each node will have to transmit a token before the actual data. The algorithm that generates the token is the core of this solution because the entire authentication is based on the validity of the token.

Both requesting and sleeping node will have to generate the same token before a communication establishes. The sleeping node will generate the token to store it at the wake up receiver and the requesting node will have to generate the same token to transmit it in order to request a remote wake up.

As we mentioned previously each token transmitted in the network has to be unique to avoid replay attacks. This requirement leads us to use a security authentication method called one-time-password. Using one time password we discourage adversaries trying to capture the tokens because the captured passwords are useless after their initial authentication.

In order to achieve a scheme that is energy efficient we avoid transmitting the generated tokens to the network because radio communication costs a lot of energy (the rest doesn't make sense as there is no explanation as to why a node needs to populate each token prior to waking up) and each node should populate by transmission its token to all its neighbours. Therefore we propose the counter-synchronised one time password for token generation. Using counter-synchronization, nodes will simply have to maintain the correct counter time and use it to generate the wake-up token as and when required. So each node uses the counter value as the token to wake-up its neighbours.

## 4.2 Encryption

Encryption of the token will have to be used for two important reasons. The first reason is to prevent adversaries from reading the token and use it to generate the next one, which consequently causes a sleep deprivation attack risk. The second reason is to differentiate the tokens for each sensor node so that the wake-up requests are always unique and do not conflict with each other and avoid overhearing of the wake up token.

We have decided to use symmetric cipher encryption because analysis has proven that it is the most suitable for lightweight applications [14-16] particularly the TEA algorithm that consumes only 7.37 µW or AES that is more effective and can be used in case we consider that TEA is not secure enough.

Combining token generation with encryption, the algorithm generation derives from the following algorithm Token = $TEA_K$(Counter-Value) where K is the unique static encryption key for each node or can also be referred to as the ID of the node.

## 4.3 Time Slots

In order to minimize the cost of computation power we need to separate the time into slots so that each token calculation will be generated to be valid for period of time and not just a moment in time. As for the Equation 2 where $t$ is the time slot, $Pw_n$ is the power needed to wake up and where $Pc_n$ is the power needed to calculate the token.

$$\sum_{t=1}^{n} t = Pw_n + Pc_n$$

Equation 2: Relationship between time slot and power consumption.

By enlarging the duration of each time slot the sum of the power consumption will be minimized dependent on the increase in duration but by doing this we limit the total number of possible authentications that can take place because for each slot only one authentication can be undertaken as illustrated in Figure 4.

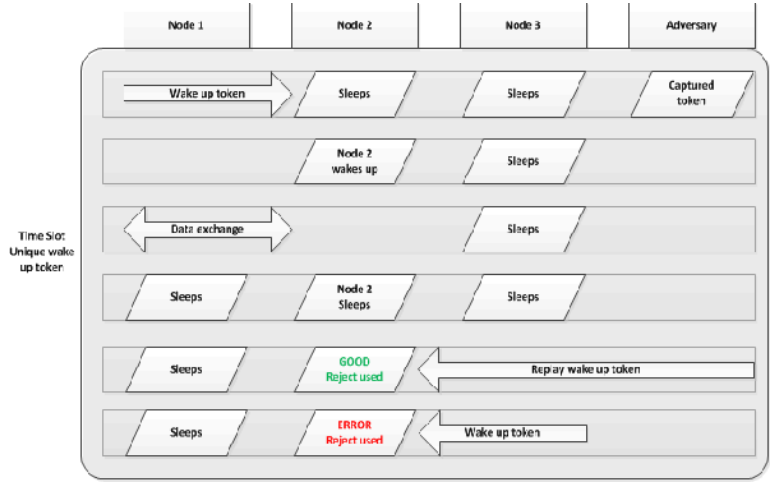

Figure 4: Our solution under replay attack.

In case a node has to receive two data streams from two different nodes in one time slot, the second node will have to be rejected to ensure that the data is fresh and not a replay of the previous data used by an adversary. The proposed solution to this problem, which also comes with a new problem attached, is to alternate the token after it has been used either by encrypting it again or by following a hash-chain every time is used [17]. However we consider this method ineffective because it can cause delays depending on the load of the network. Those delays could self-damage the requesting nodes until they realize that the original token has been already been used. We are currently working on the most efficient handle of multiple authentications on a single time slot.

### 4.4 Wake-up Receiver and costs

An example of an ultra-low wake up receiver that could be used on this scenario is the ATA5283[18] which claims that it uses only 1.2 uA to listen which means that it can listen continuously on the channel for ten years using a single AA battery, which is possibly less than the self-discharge rate of the battery when idle. Jie Wang et al [19] proposed a circuit modification to harvest enough power from the incoming signal to power the wake up receiver which also leads to complete immunization from possible attacks to the wakeup receiver itself.
Therefore by using the ATA5283 as an additional receiver, configured to always be in active mode we can achieve an energy efficient asynchronous system, capable of handling the reception of data at all times asynchronously.

### 4.4.1 Harmonic Collaboration

Although wake-up receiver and the main transceiver have different characteristics there is a requirement to ensure that the network will operate as expected. The most important requirement is that main radio and wake up radio should harmonically collaborate regarding timing and radio range. The main transceiver has to be at the same range as the wakeup receiver so that both wakeup and main radio can communicate with the same neighbours. Synchronization of the sender and the receiver regarding the time required to respond to the signal is also an important factor to be considered in order to avoid conflicts and corrupted messages. Acknowledgments can be used also if required to ensure that the transceiver is active after the wake up request has been received but sacrifices the energy resources of the requesting node, therefore this option should only be implemented where it is considered business critical, and also by exception only.

### 4.4.2 Node initialization

During the initialization, the MCU must generate an encryption key derived by its ID as defined by the data sink. This key will be used to encrypt the token so that adversaries cannot capture, identify the pattern and then generate a wake up token. In addition each node should discover its' neighbours to be able to encrypt the tokens using the remote ID. Exploration of neighbours is an unavoidable procedure that is also used by routing so that nodes will know the shortest path to the sink.

### 4.4.3 Power cost

Another problem that the scheme could face is the maintenance of the correct network time to follow the time stamps and the energy cost. For the node to maintain the correct time the MCU must remain active. Atmel also introduced a new technology called picoPower technology which allows the MCU to sleep but keep the Real time counter ticking. The outcome of this design leads to a 650nA consumption or autonomy using a single battery for over 400 years[20].

### 4.4.4 Wake-up Procedure

Each one of the nodes should have its own encryption key along with a list of its neighbour's encryption keys. For most of the time the nodes should be asleep and waiting for an interruption. The interruptions can be one of the following, a sensor interruption to take a reading, a wake-up receiver interruption that activates the node to wake it up in order to receive data by another node, or a refresh token interruption that originates from the MCU and occurs when the active token expires because the time slot has expired.

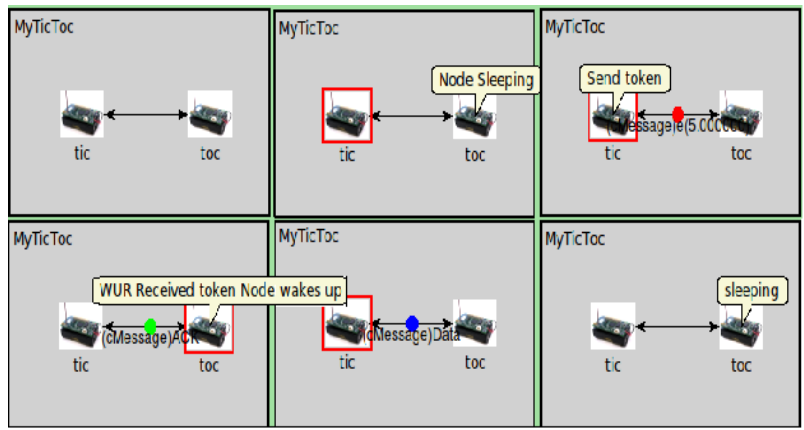

Figure 5: Simulation using omnet++ of previous proposed scheme.

Our proposed scheme advocates that the wake-up receiver never sleeps; it must remain active and wait to receive an authenticated encrypted token which is the next one in the series that is currently stored in the receiver. So in case another node "Node2" (for example) has taken a reading and needs to transmit it through this node "Node1", it must first calculate the token derived by the time slot and encrypt it using the encryption key that the destination node "Node1" will accept. This will cause the target nodes' wakeup receiver to wake the entire node and prepare the actual receiver for reception. After the transmission is complete both nodes should either continue their tasks or go back into sleep mode. In case they plan to go in to sleep mode, they must first store in their wake-up receiver the new token.

### 4.4.5 Anti-replay

One of the features of our protocol is to counter attack replay attacks. An adversary within close proximity could be listening for the wake up tokens and try to inject them into the network to engage a sleep deprivation attack. Therefore we have simulated the protocol using omnet++ to prove that a replay attack would fail to authenticate to the nodes. The network setup consists of three nodes Node0, Node1, Node2 and an adversary as shown in Figure 6. Node0 and Node2 will randomly initiate a connection to Node1 to transmit a data message but every time the adversary node receives a wakeup token it will replay (re-transmit) it to the network in an attempt to wake up the node and send malicious data to it. Figure 7 illustrates the results of our simulation filtered to show only the important information. As you can see each time the adversary receives a message it broadcasts the token and data to the nodes but the nodes do not accept the token and the data message is not received because at the main receiver remains asleep as was intended.

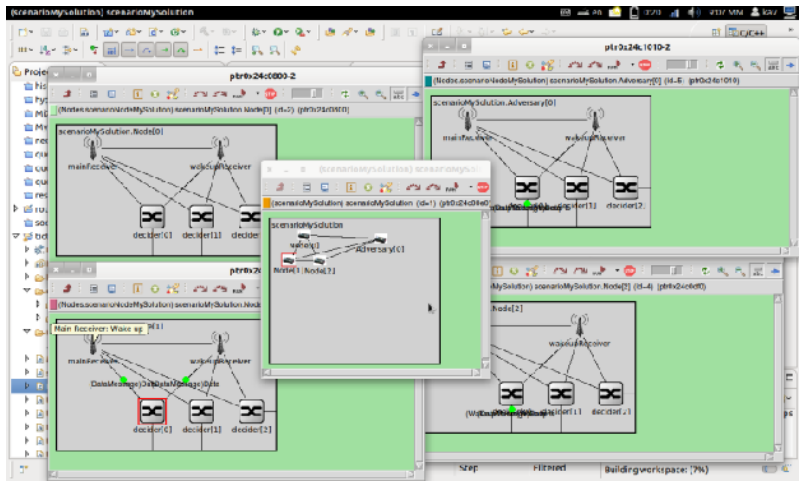

Figure 6: Simulation using omnet++ of our scheme.

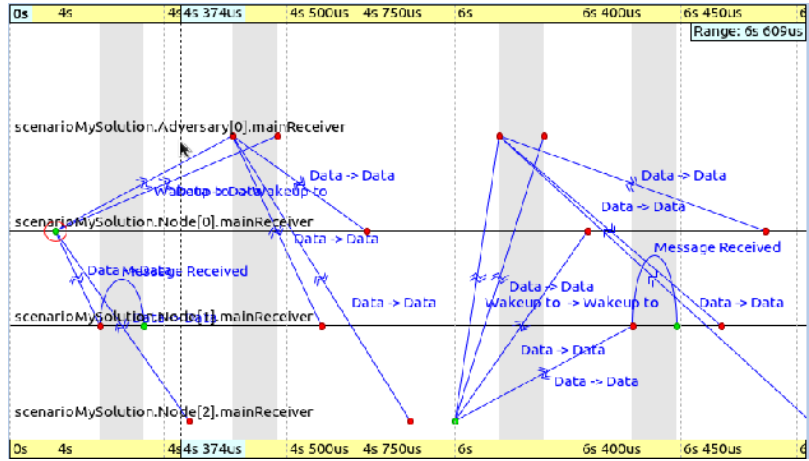

Figure 7: Graphic representation of the transmitted messages against time.

## 5. CONCLUSIONS

Proposed scheme can effectively immunize the network from deprivation attacks by moving authentication from application level to physical layer and simultaneous reduce unnecessary traffic by calculating the tokens instead of exchanging them. Calculation power is insignificant compared to transmission and reception power and at the same time does not expose cryptographic material to adversaries. We consider that our scheme under attack by denial of sleep attacks will consume less power than the self-discharge of the batteries. We are currently working on simulating the same scenario but on power aspects and solving the problem that occur on concurrent connections at the same time slot.

## AUTHORS

Salem Aljareh
Anastasios Kavoukis

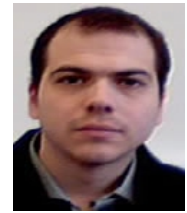